\documentclass[prl,twocolumn,floatfix,epsf,psfig]{revtex4}
\pdfoutput=1 %% this is very important command to upload pdf figures in condmat
\usepackage{epsf}
\usepackage{psfig}
\usepackage{graphicx}

\usepackage[hypertexnames=false,linktocpage=true]{hyperref}
\hypersetup{colorlinks=true,linkcolor=blue,anchorcolor=blue,citecolor=blue,menucolor=blue,filecolor=blue,urlcolor=blue}

\begin{document}
	\title{Enhancement of the density of states at the Fermi level due to oxygen atoms in noble metals}
	\author{Sudipta Roy Barman$^{1}$ and Aparna Chakrabarti$^{2,3}$ } 
	
	\affiliation{$^{1}$UGC-DAE Consortium for Scientific Research, Khandwa Road, Indore, 452001, Madhya Pradesh, India}
	\affiliation{$^{2}$Theory and Simulations Laboratory,  Raja Ramanna Centre for Advanced Technology, Indore, 452013, India}
	\affiliation{$^{3}$Homi Bhabha National Institute, Training School Complex, Anushakti Nagar, Mumbai, 400094, India}

	\begin{abstract}
		The interaction of oxygen with noble metals such as silver  has been an important topic of research for many decades. Here, we show occurrence of a peak in the density of states (DOS) at the Fermi level ($E_F$) when oxygen atoms occupy disordered  substitutional  positions in noble metals such as Ag, Au or Ag-Au alloy. This results in large enhancement of DOS at $E_F$ with respect to Ag or Au metal. Its origin is attributed to  O 2$p$ related  disorder broadened flat band that straddles almost all the high symmetry directions of the Brillouin zone. Our work suggests that if a large concentration of disordered oxygen can be realized in nano-structures of noble metals, it may lead to interesting phenomenon.     
	\end{abstract}
	%\pacs{} 
	\maketitle

\noindent{\it \bf Introduction:}\\
From way back, the interaction of oxygen with noble metals such as silver has fascinated scientists through  different intriguing phenomena  such as high solubility of oxygen in liquid Ag that changes its freezing point\cite{Baker65},  dissociation of oxygen molecule at the silver surface\cite{Gravil96}, incorporation of oxygen atoms (O-atoms) in the bulk\cite{BaoErtl92,Eberhart92,Crocombette02,Baird99}  enhancing  its catalytic activity\cite{NagySchleogl99,Li03}, %({\it \onlinecite{Santen87,NagySchleogl99}}), 
~ elongation of  free standing silver atom chain by incorporation of alternate O-atoms\cite{Thijssen06}, and anomalous superconducting proximity effect where Ag layer  increases the transition temperature ($T_C$) of Pb by 15\%\cite{Bourgeois02}.  Density functional theory (DFT) calculations by Gravil $et~al.$\cite{Gravil96}  showed  that oxygen dissociates into   chemisorbed atomic species with no barrier, while Li $et~al.$\cite{Li03} showed that O-atoms in Ag are responsible for the catalysis.  An experimental study by Bao $et~al.$\cite{BaoErtl92} reported that incorporation of O-atoms in the subsurface  causes a 3\% expansion of the topmost layer.
~Eberhart $et ~al.$ performed a first-principles calculation of the charge density to find that the transport of O-atoms  through the octahedral interstitial sites is favored in Ag and Cu, but not in Au\cite{Eberhart92}. The authors however did not consider the  vacancies, and a later DFT study showed that rather than existing as an O atom- vacancy pair, the O-atom prefers to occupy the Ag vacancy $i.e.$ exist at the substitutional site\cite{Crocombette02}. 
~Baird $et~al.$ proposed how an O-atom would diffuse in Ag  by absorbing a phonon  to overcome the potential energy barrier\cite{Baird99}. Among the different types of O-atoms that were identified in Ag, O$_{\beta}$ is the bulk dissolved oxygen, while O$_{\gamma}$ is a strongly bound species\cite{NagySchleogl99}. 
	The stoichiometric oxides of Ag are semiconducting and %such as Ag$_2$O, Ag$_2$O$_3$  and in particular  AgO 
~these have been studied by DFT %because of the similarity of Ag$^{2+}$ with Cu$^{2+}$, which is related to 
~primarily to understand why high $T_C$ superconductivity is absent in Ag oxides in spite of their similarity with the Cu oxides\cite{Pickett15,Allen11}. 

	Compared to the bulk, Ag nano-particles (NP) are more susceptible to oxidation\cite{Shankar12,Sloufova04}. Large concentrations of O and C  (0.55 and 5 times, respectively with respect to Ag) in Ag@Au core-shell particles (where Ag is the core and Au is the shell)  have been reported from x-ray photoelectron spectroscopy (XPS)\cite{Sloufova04}. Significantly, the  oxygen and carbon signals increase for XPS  measurement performed in the bulk sensitive mode, indicating that these are present primarily within the  NP and not at the surface.
	A detailed work  using x-ray absorption fine structure spectroscopy and molecular dynamics by 
	Shibata $et~al.$  showed existence of vacancies in Au@Ag  NP at room temperature that facilitates interface alloying\cite{Shibata02}.  A scanning tunneling microscopy study showed that with application of electrochemical potential, vacancy clusters can grow and diffuse on Ag-Au alloy\cite{Oppenheim91}.
	 This effect is pronounced for  small particles of size less than 4.6~nm, where the two metals are nearly randomly distributed.
		~Recently, Yue $et~al.$ have demonstrated  by transmission electron microscopy and theoretical simulation 
	that oxygen molecules dissociate on the surface of silver NP and diffuse through them to reach the silver/carbon interface and eventually oxidize the carbon\cite{Yue16}.  The lattice distortion caused by oxygen concentration gradient within the silver nanoparticles provides the direct evidence for oxygen diffusion.

In this work, we investigate how  the electronic structure of fcc noble metals such as Ag and Au is modified  by O-atoms in the disordered substitutional ($i.e.$ O-atoms in  metal vacancy positions) and interstitial  positions using the   spin polarized fully relativistic Korringa Kohn Rostoker (SPRKKR) method\cite{Ebert}. For O-atoms in the substitutional position, we find a huge increase in the density of states (DOS) at the  Fermi level ($E_F$) caused by a sharp peak related to O~2$p$-like states.  The Bloch spectral function, which is the counterpart of dispersion relation for an ordered solid, shows flat and narrow regions straddling the Fermi level along almost all the high symmetry directions of the Brillouin zone (BZ). This is similar to a flat band in an ordered solid that has been broadened by disorder. We show that this phenomenon also occurs in Au, as well as  Ag-Au alloy. We hence argue that in %non-equilibrium conditions or in 
~ nano-structures,  presence of disordered oxygen bound by dispersive force and stabilized by the noble metal matrix might %lead to enhancement of DOS at the Fermi level that might 
~lead to enhancement of conductivity.\\
% I have shifted ref 19, please check 
{\noindent \bf Methods:}\\
We have performed self-consistent band structure calculations using  SPRKKR method in the atomic sphere approximation\cite{Ebert} within the generalized gradient approximation\cite{Perdew96}. The site disorder was treated by coherent potential approximation (CPA) that calculates the configurationally averaged electronic structure self consistently within a mean-field theory. In general, CPA focuses on understanding the scattering of electrons in a material which exhibits spatial inhomogeneity and has turned out to be  the most well established way of calculating the influence of disorder on the electronic properties of different materials. 
~ The angular momentum expansion up to $l_{max}$= 4 has been used for each atom. The energy convergence criterion and CPA  tolerance has been set to 10$^{-5}$ Ry. BZ integrations were performed on a 45$\times$45$\times$45 mesh of $k$-points in the irreducible wedge of the BZ. %Spin polarized calculations were  performed with small starting moments, but the converged result did not show any moment. 
~We have also carried out spin polarized calculations  within the SPRKKR formalism. However, we arrive at converged SCF results where the systems turn out to be non-magnetic $i.e.$ the atoms do not carry any partial moment, although the  magnetic moments were  0.01$\mu_B$ for Ag and as large as 1$\mu_B$ for O in the starting configuration for all the different Ag/Au-O calculations presented here.

The supercell calculation with full potential linearized augmented plane wave method\cite{wien} has been performed with 8000 $k$ points in BZ  (256 $k$ points in the irreducible BZ). The energy cut-off is about 14.5 Ry and the tolerance for energy convergence is 10$^{-4}$ Ry. The cohesive energy ($E_{coh}$), for example for Ag$_m$O$_n$, has been calculated using  $E_{coh}$= $E_{tot}$(Ag$_m$O$_n$)$-$ $m$$\times$$E_{tot}$(Ag)$-$ $n$$\times$$E_{tot}$(O), where $E_{tot}$(Ag$_m$O$_n$), $E_{tot}$(Ag) and $E_{tot}$(O) are the total energies of the concerned Ag$_m$O$_n$ system, Ag and O atoms, respectively. It may be noted that  since it is uncertain to which components Ag$_m$O$_n$ might possibly convert to, its  stability  against the decomposition into stoichiometric oxide (Ag$_2$O, AgO, or Ag$_2$O$_3$)  and O$_2$ phases is not addressed in this work.
	
\begin{figure}[tb]
 	%%\epsfxsize=93mm
 	%%\epsffile{Layout1_1.eps} %Source: Layout1_1,D:\Barman\PROJECTS\Ag-Au\theory\analysis\guo_ver4.pxp
 	\includegraphics[width=93mm,keepaspectratio]{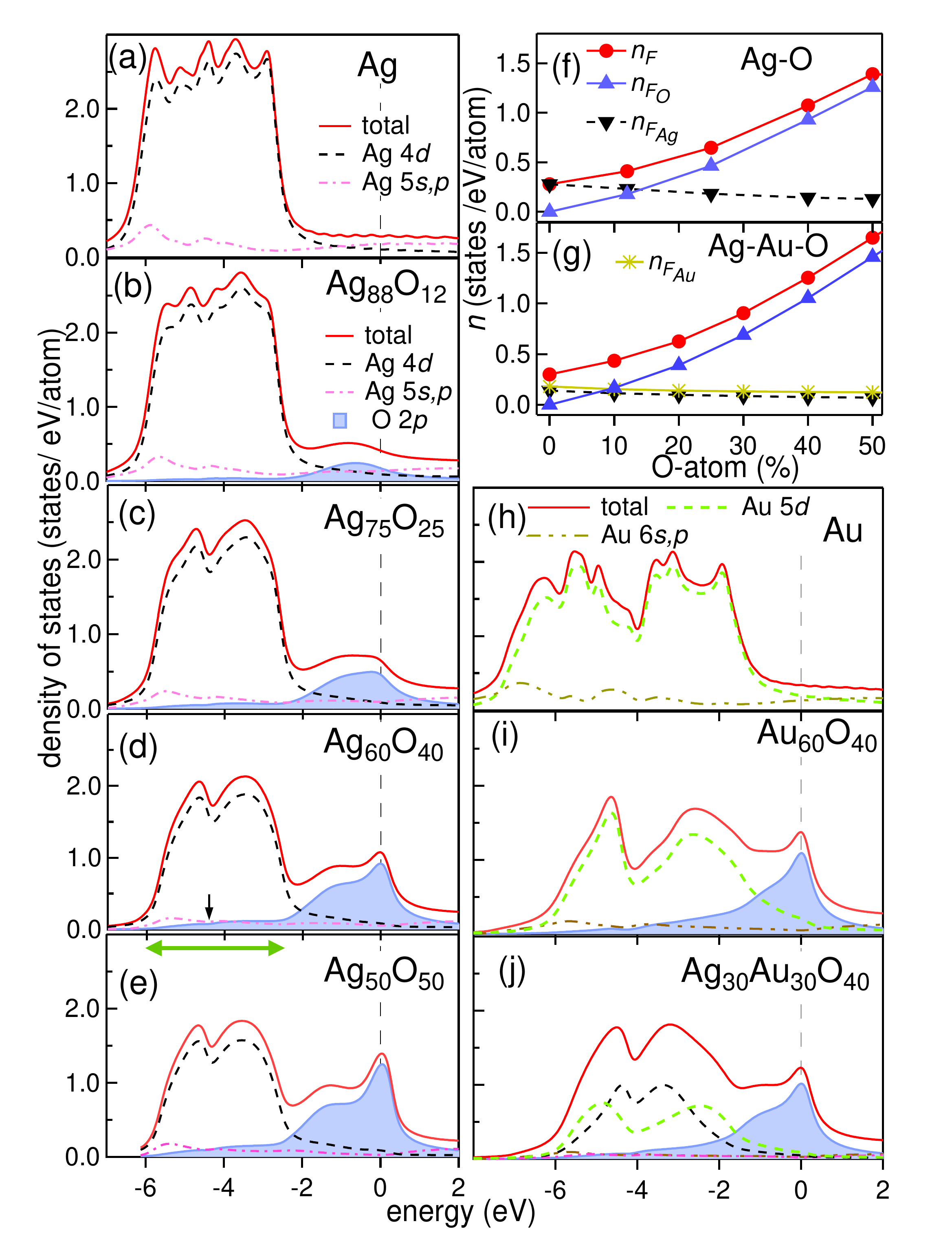} 
 	\caption{ %{\bf DOS of Ag-O, Au-O and Ag-Au-O.} 
 		Total and partial density of states of (a) Ag, (b) Ag$_{88}$O$_{12}$, (c) Ag$_{75}$O$_{25}$, (d) Ag$_{60}$O$_{40}$, (e)  Ag$_{50}$O$_{50}$, (h) Au, (i) Au$_{60}$O$_{40}$, and (j) Ag$_{30}$Au$_{30}$O$_{40}$. The total DOS at $E_F$ ($n_F$) and the total  Ag ($n_{F_{Ag}}$), O ($n_{F_{O}}$) and Au ($n_{F_{Au}}$)  contributions to $n_F$ for (f) Ag-O (g) Ag-Au-O, as a function of O-atom content.} 
 	\label{agodos}
 \end{figure}

	%  where the total energies of Ag and O $i.e.$ $E_{tot}$(Ag) and $E_{tot}$(O) has been calculated by taking a single atom  in a cubic box of size about 9\,\AA.} ~\\
~\\
{\noindent \bf Results and Discussion:}\\
  The total and  partial density of states (PDOS) of Ag with different amounts of randomly substituted O-atoms are shown in Fig.~\ref{agodos}(a-e). The calculations are performed using the equilibrium lattice constant ($a_{eq}$) obtained by fitting the variation of the total energy as a function of  lattice constant  using a % third order 
  ~polynomial function.% (Fig.~\ref{agototenergy}).
  
\begin{figure}[tb]
	%\begin{center}
%%	\epsfxsize=115mm
%%	\epsffile{Layout10.eps}
	\includegraphics[width=95mm,keepaspectratio]{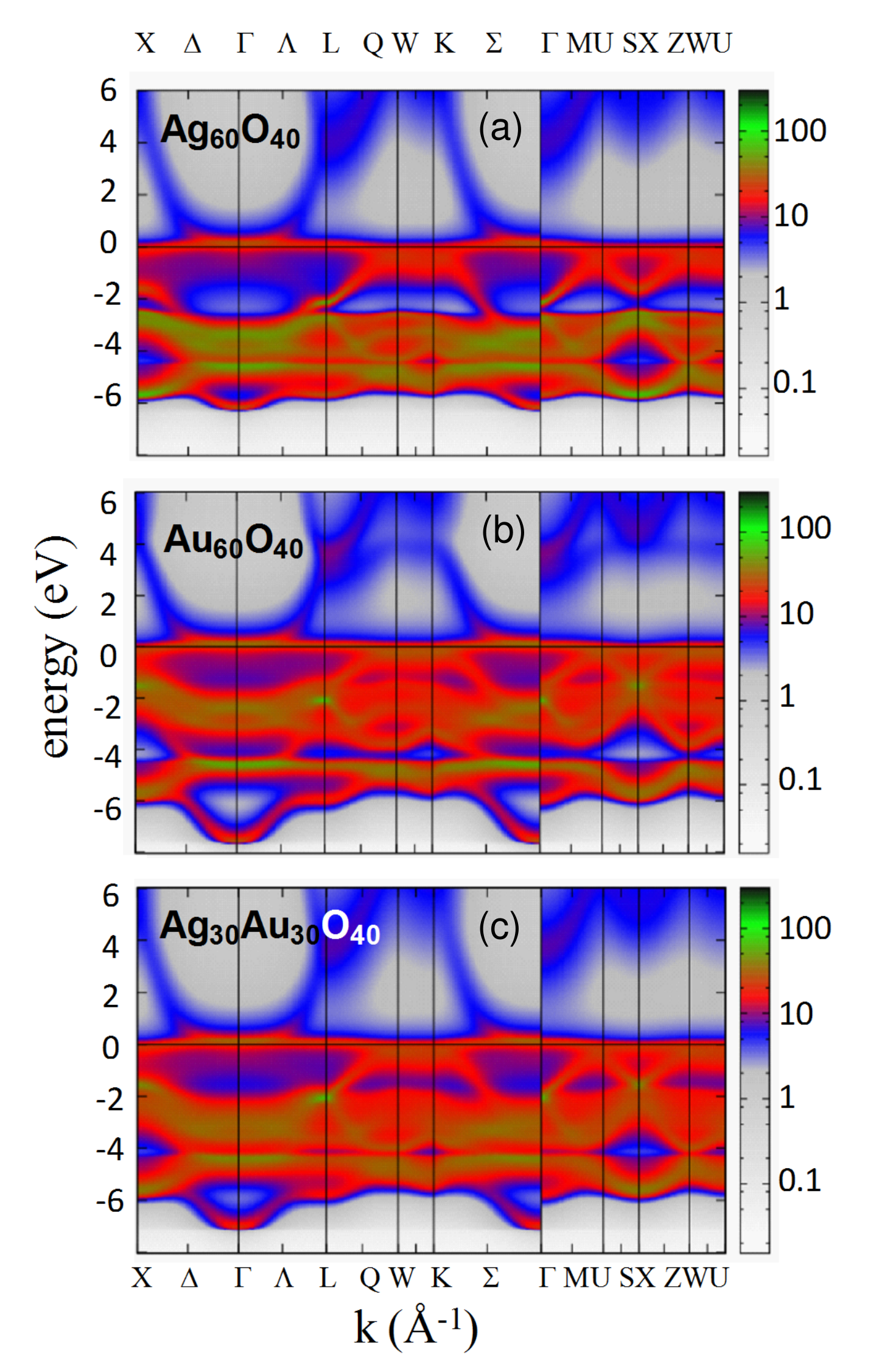} %Source: 
	\caption{ A comparison of the  Bloch spectral function $A_B$$(k, E)$ along the different high symmetry directions of the fcc BZ (a) Ag$_{60}$O$_{40}$,  (b) Au$_{60}$O$_{40}$, and (c) Ag$_{30}$Au$_{30}$O$_{40}$; the color scale on the right side shows the values of $A_B$$(k, E)$ in a.u. A high intensity of $A_B$$(k, E)$ at $E_F$ is observed in all cases resembling a disorder broadened flat band.} 
	\label{bsfAgAuO}
\end{figure}

%     \begin{figure}[tb]
     	%\begin{center}
     	%%\epsfxsize=90mm
%%     	\epsffile{Layout1_2.eps}
 %    	  	\includegraphics[width=93mm,keepaspectratio]{Layout1_2.pdf} %Source: 
  %   	\caption{ The total energy as a function of lattice constant $a$ for different Ag-O compositions, the equilibrium lattice constant $a_{eq}$ is indicated by black arrow.} 
   %  	\label{agototenergy}
    % \end{figure}
          In all cases, $a_{eq}$ turns out to be within a few percent of the Ag lattice constant (4.085\AA).  A broad hump centered around -0.7~eV in the total DOS of Ag$_{88}$O$_{12}$ that is absent in Ag  originates from the O~$2p$-like states Fig.~\ref{agodos}(a,b).   This feature causes an enhancement of the DOS at $E_F$ ($n_F$) by 46\% compared to Ag. The intensity of the hump increases in Ag$_{75}$O$_{25}$, and it shifts very close to $E_F$ (-0.2 eV). It is striking to note that, for Ag$_{60}$O$_{40}$, this O~$2p$ related peak becomes narrower and appears at $E_F$, causing a very large  increase of $n_F$ by 286\% (Fig.~\ref{agodos}(d,f)). This effect is even more pronounced for Ag$_{50}$O$_{50}$, with a 400\% increase of $n_F$, with the maximum of the O~2$p$ PDOS peak right at $E_F$ (Fig.~\ref{agodos}(e,f)). 
  %   which increases in intensity with O-atom concentration and for Ag60O40 appears right at $E_F$. 
  ~The enhancement of $n_F$  is also obtained for Au-O (Fig.~\ref{agodos}(h,i)), as well as in Ag-Au-O (Fig.~\ref{agodos}(g,j)). In fact, for the same oxygen content, $n_F$ is largest in Au-O ($e.g.$  $n_F$= 1.37 states/eV/atom for Au$_{60}$O$_{40}$ in Fig.~\ref{agodos}(i)), compared to $n_F$= 1.25 states/eV/atom for Au$_{30}$Ag$_{30}$O$_{40}$ in Fig.~\ref{agodos}(j) and $n_F$= 1.07 states/eV/atom for Ag$_{60}$O$_{40}$ in Fig.~\ref{agodos}(d)). 
 \begin{figure}[tb]
	%\begin{center}
%%	\epsfxsize=115mm
%%	\epsffile{Layout9.eps} %Source: Layout9,D:\Barman\PROJECTS\Ag-Au\theory\analysis\guo_ver4.pxp; PICT12 in Layout 9 is copy pasted from Slide4 of D:\Barman\PROJECTS\Ag-Au\theory\analysis\figure_ppt
	\includegraphics[width=95mm,keepaspectratio]{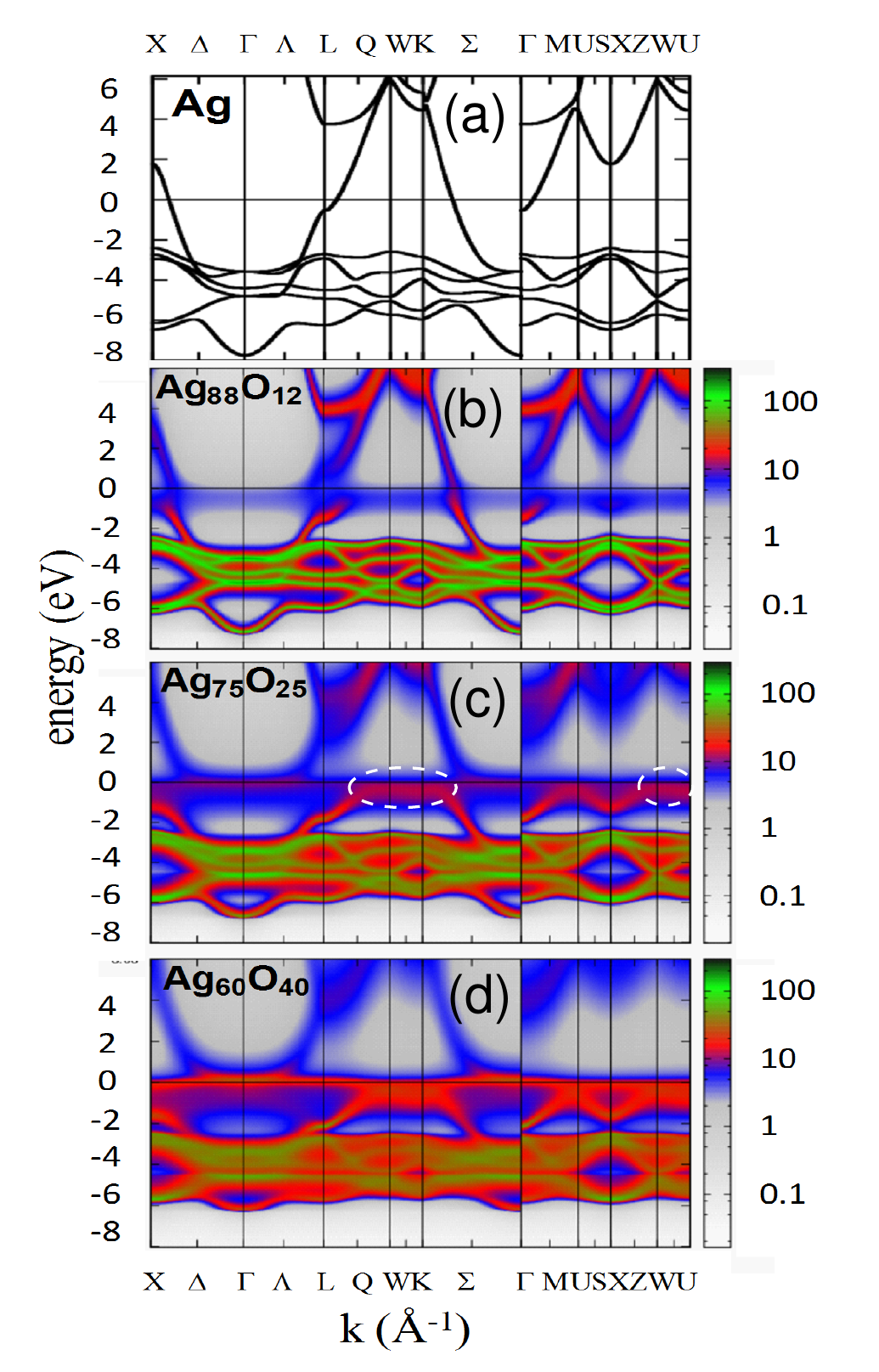} 
	\caption{%{\bf Bloch spectral function of Ag-O.} 
		(a) The band dispersion $E(k)$ along the different high symmetry directions of Ag  compared with the Bloch spectral function $A_B$$(k, E)$ along the same directions for (b) Ag$_{75}$O$_{25}$, and (c) Ag$_{60}$O$_{40}$, the color scale on the right side shows the values of $A_B$$(k, E)$ in a.u.} % A black horizontal line  at zero is the Fermi level.} 
	\label{agobsf}
\end{figure}
For a disordered system, an equivalent  of the band dispersion $E(k)$  can be obtained through  the Bloch spectral function, $A_B$$(k,E)$, which is defined as the Fourier transform of the retarded single electron Green's function\cite{Ebert11}. For an ordered system, Bloch spectral function (BSF) reduces to a set of $\delta$ functions at the band energies giving $E(k)$. With introduction of disorder,% through substitutional  impurities, 
~these $\delta$-function peaks are broadened.% and shifted. %$A_B$$(k,E)$ as function of $k$ at $E_F$ can be taken as an alloy Fermi surface. Tried to calculate, plot was showing meaningless image, should try again 
~ To find the origin of the DOS peak at $E_F$ in Fig.~\ref{agodos}, we have calculated the BSF along the high symmetry directions of the fcc BZ. If the BSF of Ag$_{88}$O$_{12}$ in Fig.~\ref{agobsf}(b) is compared to $E(k)$ of Ag (Fig.~\ref{agobsf}(a)),  it is observed that the dispersing Ag $s$,\,$p$ band with a kink at the $L$ point splits in the former. The splitting happens because the occupied band shifts down in energy to about -1.5 eV, while the unoccupied part up to 1~eV diminishes in intensity (in Fig.~\ref{agobsf}(b) see the red colored band around $L$ point, red is about 25 in the relative intensity logarithmic color scale with  green as maximum ($\approx$100) and white as minimum (zero))). The same happens  for the other dispersing Ag $s$-\,$p$ bands, $e.g.$  the bands crossing $E_F$ at middle of $\Delta$-$X$ and near $M$ point.  Thus, evidently, the contribution of Ag $s$-\,$p$ states to $n_F$ decreases. % and the Fermi surface is distorted. **try to get the FS still**  
~A curious observation from Fig.~\ref{agobsf}(b) is the appearance of states of low intensity at $E_F$ (blue region, 5 in the intensity scale) and we examine how these evolve with larger O-atom content. Strikingly, for Ag$_{75}$O$_{25}$ in  Fig.~\ref{agobsf}(c),   parts of the blue region become intense and appear as flat red bands along $W$-$U$ and  $Q$-$\Sigma$ (both encircled by white dashes). The flat band along $Q$-$\Sigma$ connects  two dispersing bands crossing $L$ and $\Sigma$. A parabolic band emerges with minimum at $X$ at -1.5~eV and disperses up to the flat band along $W$-$U$. Weak flat bands (light red color) at $E_F$ is also observed along $\Delta$-$\Gamma$-$\Lambda$ and $\Sigma$-$\Gamma$. A spectacular  effect occurs for Ag$_{60}$O$_{40}$ (Fig.~\ref{agobsf}(d)): a flat narrow  band (red color)  appears  at $E_F$ spanning  all the high symmetry directions from $X$-$\Gamma$-$L$-$W$-$K$-$\Gamma$ and $\Gamma$-$U$-$X$-$W$-$U$ that we have calculated (encircled by yellow dashes).  The very wide $k$ range of the flat band, as if it is ubiquitous, along the different directions of the fcc BZ is completely unique. Obviously, this is the origin of the sharp DOS peak at $E_F$. The intensity of this flat band  further increases  for Ag$_{50}$O$_{50}$. The BSF curves of Au$_{60}$O$_{40}$ and Au$_{30}$Ag$_{30}$O$_{40}$ also show  presence of this intense flat band at $E_F$ (Fig.\ref{bsfAgAuO}). 
 
 %Thus, in analogy with an ordered solid where flat bands give rise to van Hove singularity,   presence of flat bands obtained from the BSF  establish existence of van Hove singularity at $E_F$ in Ag/Au-O (Fig.~\ref{agobsf}, \ref{bsfAgAuO}), albeit broadened by disorder.  van Hove singularity at $E_F$  is one of the crucial parameters that determines the superconducting energy gap and  $T_C$ of a superconductor\cite{Tsuei90,Li09,Cao18} and observation of such an effect could open up exciting possibility of  superconductivity in this system. 
 
 In order to further understand why presence of O-atoms in Ag or Au causes such a large enhancement of the DOS at $E_F$ , we find from the integration of the PDOS up to $E_F$ that the total valence charge in O-atom varies from 4.4 to 4.6 $i.e.$ its valency being -0.4 to -0.6. Thus, the O-atom is almost neutral, with a small excess negative charge (2$p^4$ is the valence configuration of oxygen).  Its valency  is remarkably smaller than the nominal valency of -2 for oxygen, for example,    in stoichiometric oxides, where the covalent bonding is predominant and Ag to O nearest neighbor ($nn$) distance is $\approx$2\AA\,\cite{Allen11}. On the other hand, in Ag/Au-O systems studied here, the $nn$  distance between two O-atoms is substantially larger ($e.g.$  2.84\,\AA~ for  $a_{eq}$= 4.02\AA) and thus orbital overlap required for covalent bonding is diminished.%page 46
~Rather, it is interesting to note that the  van der Waals (vdW) diameter of the O-atom, $i.e.$ the distance between two O-atoms where the minimum of the Lennard-Jones interaction potential occurs, is reported from different estimates to be in the range of 2.8\,\AA~ to  3\,\AA\,\cite{Pauling,Bondi64}.  Thus, based on the observations that the O-atoms are almost neutral and at a distance equivalent to their  vdW diameter,  the dominance of  vdW bonding between O-atoms is expected. %% In an earlier work\cite{Schmid06}, the importance of  vdW interaction in influencing the  $p$(4$\times$4)-O surface reconstruction on Ag, rather than formation of Ag oxide, was demonstrated.
~%vdW interaction in materials and molecules is ubiquitous\cite{Parsegian05} and its 

\begin{figure}[tb]
		%\begin{center}
%%\epsfxsize=100mm
%%\epsffile{Layout11.eps}
	\includegraphics[width=93mm,keepaspectratio]{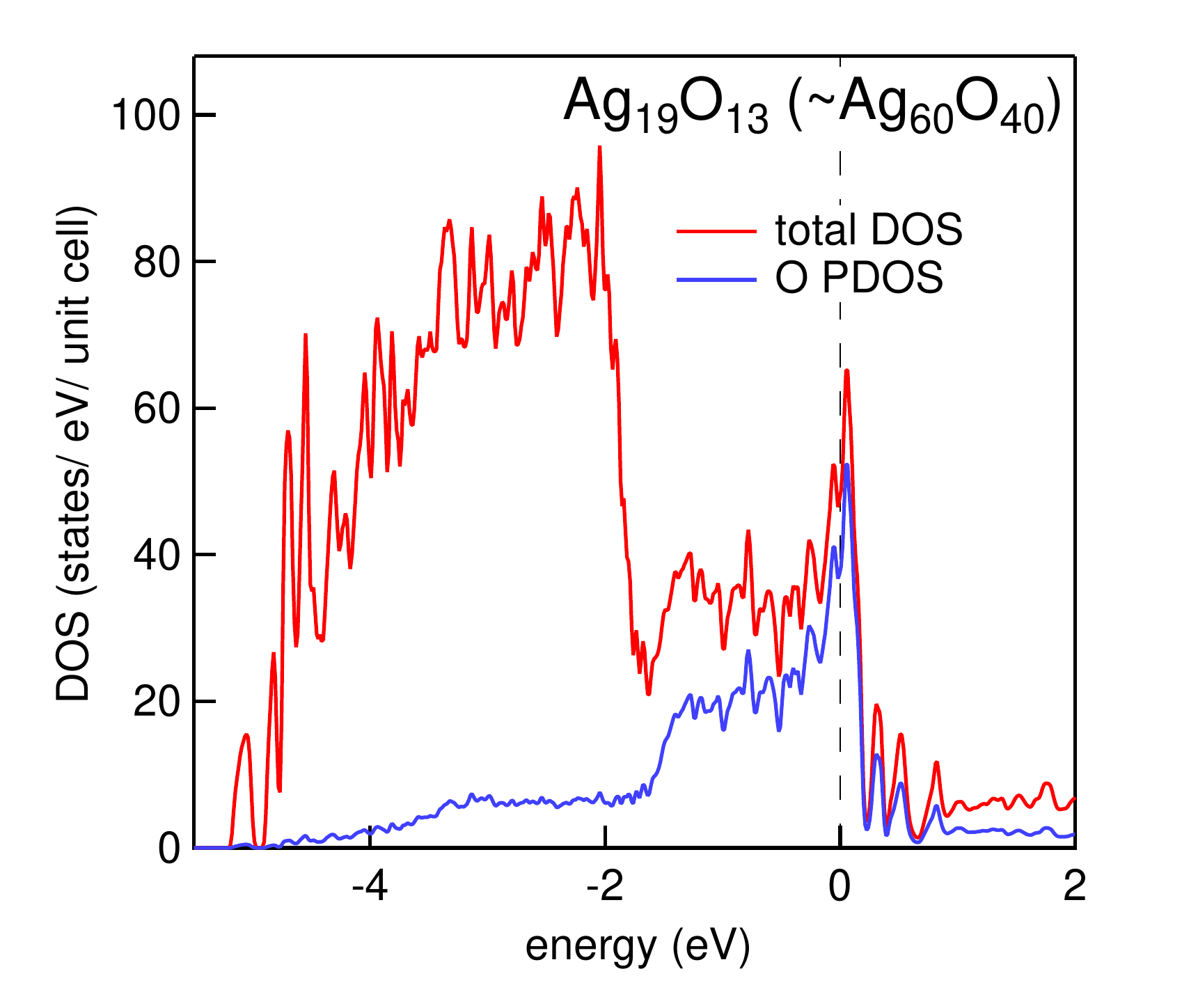} %Source: 
		\caption{The total and oxygen density of states for Ag$_{19}$O$_{13}$ calculated with a 32 atom fcc supercell using full potential linearized augmented plane wave method. } 
		\label{AgO_wien}
		\end{figure}

       In Fig.~\ref{agodos}, the O~2$p$ states extend from $E_F$ to -6~eV; it has a shoulder at -1.3~eV that  arises mainly from the 2$p_{x,y}$  states, while the peak at $E_F$ has similar contributions from all the three $p$ components. Its noteworthy that in the -2.5 to -6~eV region ($e.g.$ indicated by green horizontal arrow in Fig.~\ref{agodos}(d)), the O~2$p$ PDOS tails out as almost flat with apparently low intensity, but in reality it constitutes about 20\% of the total occupied PDOS. %p. 46
~ It has a similarity of shape with the Ag $d$ states, with both exhibiting a minima at -4.5 eV (black arrow). Such tailing of the O~2$p$ PDOS is also observed for Au$_{60}$O$_{40}$ and Ag$_{30}$Au$_{30}$O$_{40}$ (Fig.~\ref{agodos}(i,j)),  indicating  hybridization of the O~2$p$ states with the noble metal $d$ states. The cohesive energies ($E_{coh}$) of Ag$_{60}$O$_{40}$ and Au$_{60}$O$_{40}$ have been calculated by considering a 2$\times$2$\times$2 fcc supercell of 32 atoms with  random configurations of 13 O-atoms using full potential linearized augmented plane wave (FPLAPW) method\cite{wien}. The  DOS shows the O~2$p$ related sharp peak at $E_F$  (Fig.~\ref{AgO_wien}),  which is  of similar intensity as in Fig.~\ref{agodos}(d). The PDOS also has the hump around -1.3 eV and the flat DOS region. However, the DOS in this case is highly structured; this is because here we have considered only one particular configuration of O-atoms that is repeated in space in the FPLAPW method and thus it is not a fully disordered structure. The values of  $E_{coh}$ for Ag$_{60}$O$_{40}$ and Au$_{60}$O$_{40}$ are 2.8 eV/atom and 2.9 eV/atom, respectively. In comparison,  $E_{coh}$ for O in fcc structure with same lattice constant is much smaller (1.3 eV/atom). %Ag is 2.7 eV/atom and that of
~Thus,  hybridization of O~2$p$ and the Ag/Au $d$ states, as well as their cohesive energies indicate the importance of the noble metal matrix in providing the stability.
	
\begin{figure}[tb]
	%\begin{center}
%%	\epsfxsize=100mm
%%	\epsffile{Layout4_2.eps}
	\includegraphics[width=93mm,keepaspectratio]{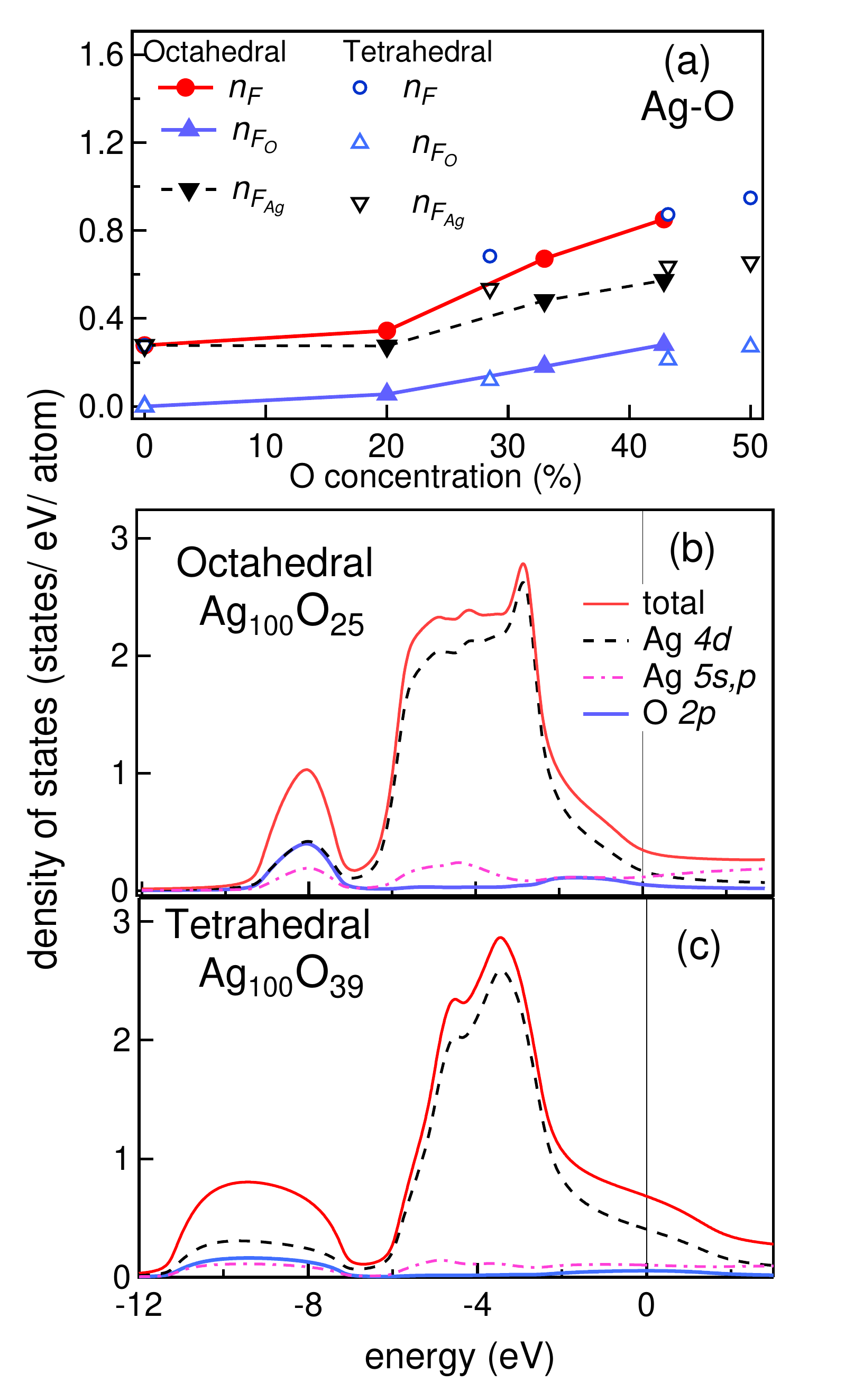} %Source: 
	\caption{ (a) The total DOS at $E_F$ ($n_F$) and the total  Ag ($n_{F_{Ag}}$) and O ($n_{F_{O}}$) contributions to $n_F$ for O-atoms in octahedral and tetrahedral interstitial positions, as a function of O-atom concentration. Total density of states, Ag 4$d$, Ag 5$s,p$  and O 2$p$ PDOS for (b) octahedral Ag$_{100}$O$_{25}$ and (c) tetrahedral Ag$_{100}$O$_{39}$, where disordered O-atoms and vacancies are considered in the interstitial sites.} 
	\label{agooctados}
\end{figure}

We have also considered O-atoms to be randomly distributed in octahedral  and tetrahedral  interstitial positions of Ag (Fig.~\ref{agooctados}(b,c)), since these have been reported in literature as possible sites that oxygen could occupy\cite{Baird99,Eberhart92,Crocombette02}.
~In both the interstitial positions, due to the close proximity of O and Ag atoms (2.04\,\AA~ for octahedral and 1.77\,\AA~ for tetrahedral sites), the Ag~4$d$-\,O~2$p$ hybridization dominates and the Ag~4$d$ PDOS is largely modified. The O~2$p$ PDOS exists over a wide energy range of 10-12 eV, and does not have any peak at $E_F$. Nevertheless, $n_F$ increases moderately with O-atom content (Fig.~\ref{agooctados}(a)). In contrast to Ag-O (Fig.~\ref{agodos}), this is caused primarily by  Ag~4$d$ and 5$s,p$ states rather than O~2$p$ states. 

\begin{figure}[tb]
	%\begin{center}
	%%	\epsfxsize=110mm
	%%	\epsffile{Layout7.eps}
	\includegraphics[width=85mm,keepaspectratio]{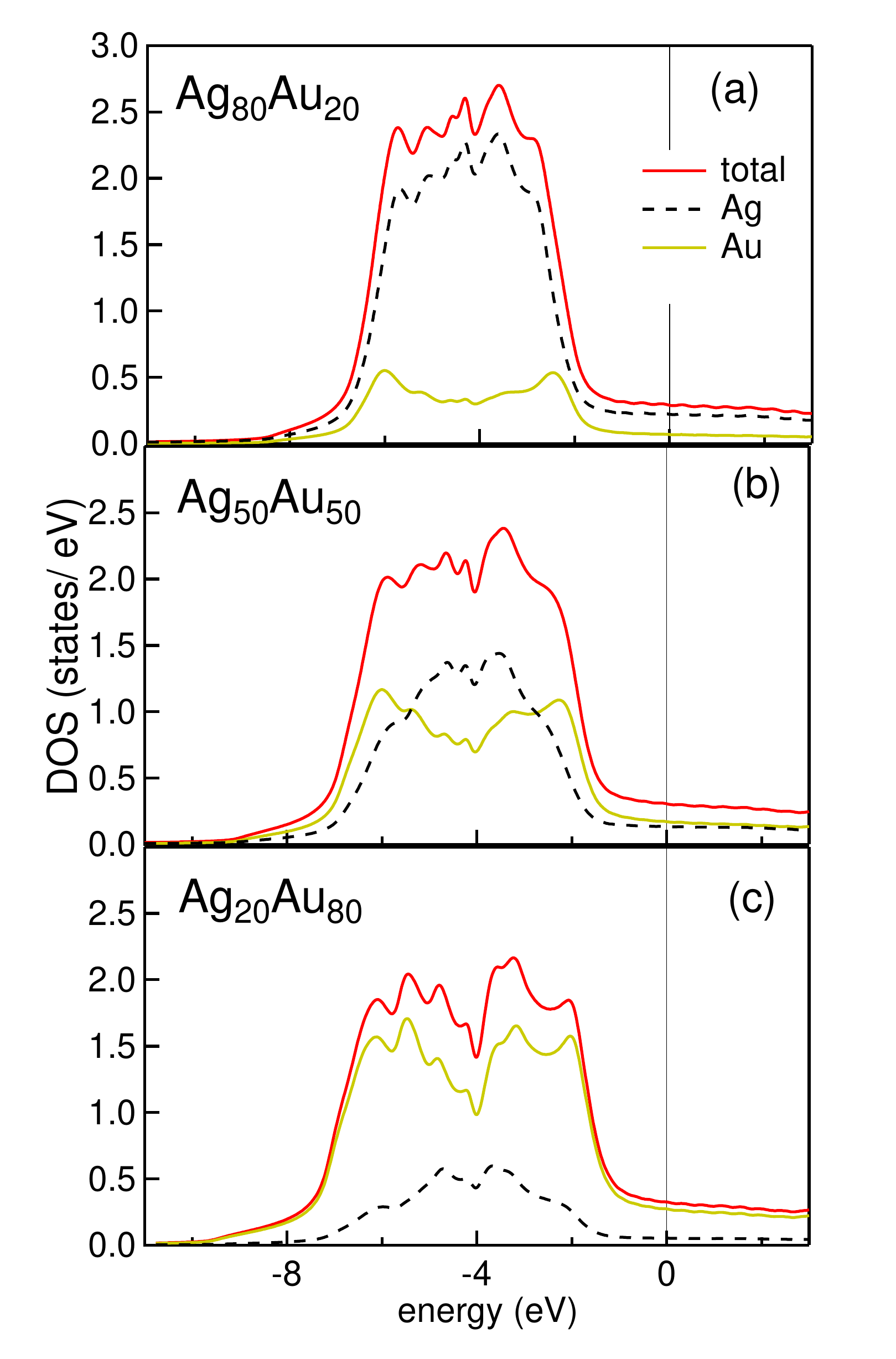} %Source: 
	\caption{The total, Ag and Au  density of states for (a) Ag$_{80}$Au$_{20}$, (b) Ag$_{50}$Au$_{50}$  and (c) Ag$_{20}$Au$_{80}$.  The position of the Ag 4$d$ and Au 5$d$ states remain essentially unchanged and their relative intensities  vary proportionately with composition, with hardly any modification in the DOS of the $s,p$ states at $E_F$.} 
	\label{agaudos}
\end{figure}

 A recent work claimed occurrence of  room temperature superconductivity  in 10\AA~ Ag NP in Au matrix\cite{Pandey18},
although it must be mentioned at the onset that reproducibility of the data has not been established in the literature yet\cite{otherarchives}. It is well-known that high temperature superconductivity could be achieved in materials with large $n_F$ and high Debye temperature, however, claim of superconductivity in pure Ag@Au nano-structure (NS) is very curious and even seems unrealistic, since neither Ag nor Au are superconducting.
 In Ref.~\onlinecite{Pandey18}, the authors did not discuss the amount of oxygen  present in the Ag@Au NS, although their EDAX data  show presence of C and O K$\alpha$ peaks at 0.28 and 0.53 keV, respectively. It is well known in literature that Ag NP are prone to oxidation\cite{Shankar12,Sloufova04} and the following estimate provides an idea about possible oxygen content: in a 10\,\AA~ Ag NP, there are about 30  atoms. %(atomic diameter= 3.2\,\AA). 
 ~If a single O-atom layer is considered to fully cover this spherical  NP, an estimate of the number of O-atoms considering these as hard spheres  with atomic diameter of 1\,\AA~  will involve 400 O-atoms. The x-ray diffraction pattern  shows that the Ag@Au NS has fcc structure with no extra phases\cite{Pandey18}. This shows that the oxygen present in the NS does not form any of the known oxides of Ag since their structures are not fcc. This indicates that the O-atoms in the NS  are present  in disordered positions that keep the fcc structure unaffected. 
%Importantly, although from bulk calculation, Ag/Au-O has lower cohesive energy compared to their stoichiometric  oxides, possibly this metastable structure  is stabilized in Ag@Au NS. 
 ~ We further point out that the O-atoms  could be present in the  Ag vacancy ($i.e.$ substitutional) sites from the evidences provided by Shibata $et~al.$\cite{Shibata02}, Oppenheim $et~al.$\cite{Oppenheim91} and Yue $et~al.$\cite{Yue16}, who establish presence of vacancies and O-atoms, their migration through the NP and complete alloying, particularly in small NP of size less than 46\,\AA. %,~  complete alloying of Ag and Au and migration of O-atoms through NP occur due to presence of vacancies 
~Moreover, a DFT study by Crocombette $et~al.$ established that it is energetically favored for an oxygen–vacancy pair in Ag to transform to a substitutional O-atom\cite{Crocombette02}. On the basis of the above arguments, it is conceivable that there is a probability that O-atoms can form local regions of disordered Ag/Au-O with high oxygen concentration encompassing the Ag core and Au matrix, thus possibly forming a connected  pathway of regions of high DOS at the Fermi level resulting in high conductivity. While the large DOS at the Fermi level due to occurrence of disordered broadened flat bands as observed in Figs.~\ref{agodos}-\ref{bsfAgAuO} may favor superconductivity\cite{Tsuei90,Li09,Cao18}, whether indeed the local oxygen induced superconducting regions may be formed in this kind of systems  needs to be probed experimentally. %which may establish the formation of these connected pathways leading to interesting phenomena, including superconductivity. 

%The quantum confinement effect that is well known for NP might further enhance $n_F$; as has been shown for Ag NP of size less than 15\,\AA~ by DFT\cite{Kiss11}.
 %However, this might be somewhat reduced in Ag@Au NS since the interface might be somewhat  blurred due to alloying. 
%The vdW bonding might be further enhanced in Ag@Au NS %
%~because it was shown from DFT that vdW bonding reduces their formation energy\cite{Miotto14}. %Moreover, potential role of vdW interaction in influencing the  p(4x4)-O surface reconstruction on Ag (rather than formation of Ag oxide) has been pointed out ({\it \onlinecite{Schmid06}}).
%~In an interesting work  for sulfur protected Au NP,   formation of Au-Au bonds that reduce the covalent contribution to Au-S bonds was demonstrated theoretically\cite{Reimers16}.% While calculation of phonon density of states in this disordered system is at present not possible, we  
%~Another important point is that O-atoms have relatively large volume available in the substitutional position because of their smaller size. % than the Ag atom. %the atomic diameters of Ag and O are 3.2\AA~ and 1\AA,~ respectively. This can give rise to two possible interesting scenarios.
%~This could possibly lead to formation of polarons and/or  ratting phonon modes, both of which are  known to favor superconductivity. %enhance $T_C$, $e.g.$ in pyrochlore oxide superconductor.
\begin{figure}[tb]
	%\begin{center}
%%	\epsfxsize=120mm
%%	\epsffile{Layout8.eps}
	\includegraphics[width=93mm,keepaspectratio]{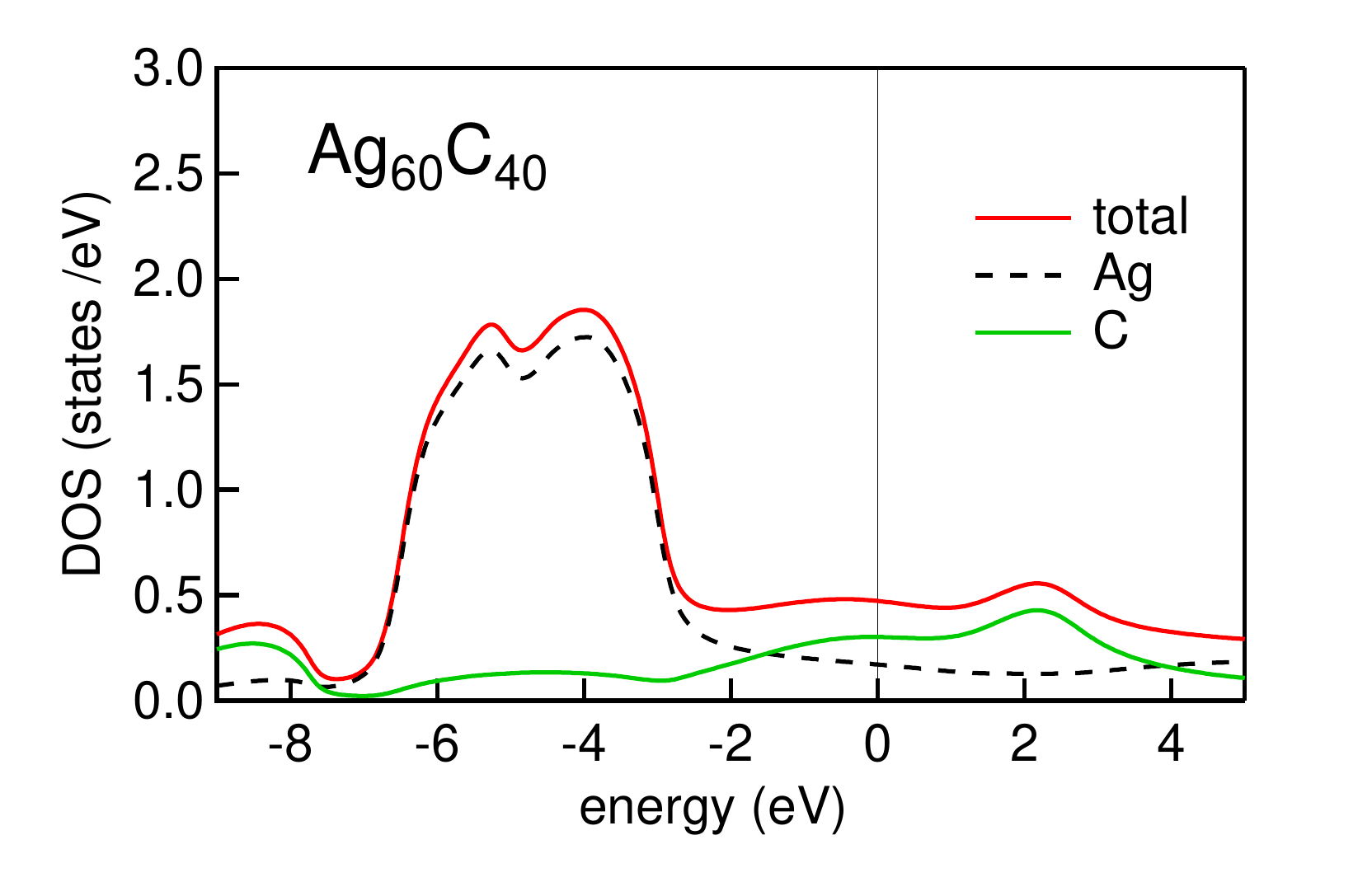} %Source: 
	\caption{(a)~The total, Ag and C density of states for  Ag$_{60}$C$_{40}$, where C is in disordered substitutional positions. The calculation has been performed using optimized lattice constant  (3.922\,\AA). %7.41256 au,
		~ A  peak in the C 2$p$ PDOS  appears above $E_F$ at 2.2 eV.} 
	\label{agcdos}
\end{figure}

We also probe the possibility of alloying of Ag with Au in an Ag-Au alloy. In Fig.~\ref{agaudos}, we show the DOS of site-disordered Ag-Au alloys with composition varying from Ag$_{80}$Au$_{20}$ to Ag$_{20}$Au$_{80}$. Neither any enhancement of $n_F$ nor transfer of charge from Ag to Au is observed from our SPRKKR calculation. We have also considered disordered carbon in Ag (Ag-C), as carbon has been observed to be present in the EDAX signal of Ag@Au NS\cite{Pandey18}. The DOS of Ag$_{60}$C$_{40}$ in Fig.~\ref{agcdos} calculated using optimized lattice constant of 3.922\,\AA~ %7.41256 au,
~does not show any peak at $E_F$.  $n_F$ turns out to be 0.47 states/eV/atom, which although is larger than Ag, is considerably  less compared to  Ag$_{60}$O$_{40}$. \\
~~\\
{\noindent \bf Conclusion:}\\
We show from density functional theory calculations using SPRKKR method that incorporation of oxygen atoms in disordered substitutional positions of noble metals such as Ag, Au and Ag-Au alloy leads to a large enhancement of the density of states  at the Fermi level. The Bloch spectral function, which is the counterpart of dispersion relation for an ordered solid, shows that the peak in the density of states at the Fermi level  is related to  disorder broadened flat O 2$p$ states that straddle almost all the high symmetry directions of the Brillouin zone.  We argue that large concentration of disordered oxygen atoms, if can be realized in noble metal nano-structures, may result in interesting phenomena and stimulate new experiments on noble metal nano-structures. %, the first step would be to microscopically investigate the distribution of the O-atoms.  

%and formation of continuous possibly superconducting regions of  Ag-Au-O between the probe leads at some step(s) of the NS fabrication that involves complicated chemical reactions is probably uncontrolled at present. This is presumably a reason for very large variation of $T_C$ (140 to 320~K) in Ref.\,\cite{Pandey18}.% and inability  of other laboratories to reproduce their data.

 %even  if the claim in Ref.\,{\it \onlinecite{Pandey18}} turns out to be non-reproducible, 
%The novelty of our work is to point out the prospect of superconductivity related to van Hove singularity at the Fermi level %in a bulk or nano-structured system 
%~if van der Waals bonded oxygen can be stabilized in a noble metal matrix.  
{\noindent \bf Acknowledgments:}\\
We  thank the Computer Centre of Raja Ramanna Centre for Advanced Technology, Indore  for providing the computational facility. A.C. thanks P.A. Naik for support and encouragement.\\

\end{document}